\documentclass[12pt]{iopart}
\usepackage{bm}
\usepackage{hyperref}
\newcommand{\dn}[2]{{\mathrm{d}^{{#1}}{{#2}}}}
\newcommand{\measure}[1]{{\mathrm{d}{{#1}}}}
\newcommand{\fmeasure}[1]{{[\mathrm{d}{{#1}}]}}
\newcommand{\deriv}[2]{{\frac{\mathrm{d}{{#1}}}{\mathrm{d}{{#2}}}}}
\newcommand{\pd}[2]{{\frac{\partial{{#1}}}{\partial{{#2}}}}}
\newcommand{\grad}{\nabla}
\newcommand{\R}{\mathcal{R}}
\newcommand{\cs}{c_{\mathrm{s}}}
\newcommand{\vect}[1]{{\bm{\mathrm{{#1}}}}}
\newcommand{\planckmass}{M_{\mathrm{P}}}
\renewcommand{\e}[1]{{\mathrm{e}^{{#1}}}}
\newcommand{\imag}{{\mathrm{i}}}
\newcommand{\ket}[1]{{|{{#1}}\rangle}}
\newcommand{\bra}[1]{{\langle{{#1}}|}}
\newcommand{\braket}[2]{{\langle{{#1}}|{{#2}}\rangle}}
\newcommand{\fnl}{f_{\mathrm{NL}}}
\renewcommand{\epsilon}{\varepsilon}
\begin{document}
\title{Primordial non-gaussianities in single field 
inflation}
\date{\today}
\author{David Seery and James E. Lidsey}
\address{Astronomy Unit, School of Mathematical Sciences\\
  Queen Mary, University of London\\
  Mile End Road, London E1 4NS\\
  United Kingdom}
\eads{\mailto{D.Seery@qmul.ac.uk}, \mailto{J.E.Lidsey@qmul.ac.uk}}
\submitto{JCAP}
\begin{abstract}
We calculate the three-point function for primordial scalar fluctuations 
in a single field inflationary scenario where the scalar 
field Lagrangian is a completely general function of the field and 
its first derivative.  We obtain an explicit expression for the three-point 
correlation function in a self-consistent approximation scheme where the
expansion rate varies slowly, 
analogous to the slow-roll limit in standard, single-field inflation.  
The three-point function can be written in terms of the familiar slow-roll 
paramters and three new parameters which measure the 
non-trivial kinetic structure of the scalar field, 
the departure of the sound speed 
from the speed of light, and the rate of change of the sound speed.
\vspace{3mm}
\begin{flushleft} \textbf{Keywords}: Cosmological perturbation theory,
Inflation,
Cosmology of theories beyond the SM,
Physics of the early universe\\[2mm]
astro-ph/0503692
\end{flushleft}
\end{abstract}

\maketitle
\section{Introduction}
In the inflationary scenario driven by a single self-interacting 
scalar field, the dominant mode of primordial 
fluctuations is predicted to be Gaussian to a very good 
approximation. However, it has long been appreciated
that non-trivial information should be encoded in the connected three-point 
function and higher connected correlation functions
\cite{allen-grinstein,falk-rangarajan,gangui-lucchin,acquaviva-bartolo,%
verde-trispectrum}. 
These correlation functions are expected to leave signatures in 
the statistical properties of the cosmic microwave background (CMB) 
temperature anisotropies, and may already have been detected 
\cite{mcewen-hobson,mukherjee-wang,vielva-martinez-gonzalez}.  
In principle, such information yields crucial 
insight into the nature of the underlying
scalar field Lagrangian during the inflationary epoch, 
and provides a sensitive discriminant 
\cite{wang-kamionkowski,babich-creminelli,komatsu-nongaussian} 
between the large number of competing inflationary models 
\cite{alishahiha-silverstein,arkani-hamed-creminelli,creminelli,gruzinov,%
gupta-berera}.

In the standard single field inflationary scenario, 
the scalar field action is generally taken to be of the form
\begin{equation}
  \label{intro:scalaraction}
  S = - \int \dn{4}{x}\; \sqrt{-g} 
  \left( \frac{1}{2}\left( \partial \phi \right)^2 + V(\phi) \right) ,
\end{equation}
where the kinetic term is canonically normalized. 
Choosing a suitable inflationary model then corresponds to 
engineering an appropriate form for the potential, $V(\phi)$.  
Unfortunately, identifying an acceptable form for the inflationary potential
has proved to be a difficult task \cite{lyth-riotto}.  

The scalar field potential is not necessarily the 
only degree of freedom in inflationary model building. 
Indeed, in models descending from a supergravity or 
superstring compactification, where the inflaton might be 
identified with a light moduli field, it is 
generally expected that corrections to the kinetic term 
of the scalar field action (\ref{intro:scalaraction}) 
will arise \cite{dewolfe-giddings}.
Moreover, even if the description of microphysical degrees of freedom 
given by the action \eref{intro:scalaraction} is appropriate 
at the classical level, one would generally expect 
loops in the quantum theory to generate operators in the Lagrangian 
that are proportional to higher derivatives
$(\partial\phi)^2$, $(\partial\phi)^4$, and so on. Such interactions would be
suppressed by powers of the renormalization scale $M$, and   
if this scale is large, $M \sim \planckmass$, 
where $\planckmass$ is the Planck mass, 
the contribution of these operators would be 
negligible at the energy scale of inflation.  
On the other hand, if $M$ is closer to the unification scale 
of some Grand Unified Theory,
ultraviolet corrections of this type might be significant 
and of considerable relevance in the very early universe 
\cite{armendariz-picon,garriga-mukhanov}.

Such non-minimal choices of the scalar field action can be written in the form
\begin{equation}
  \label{intro:noncanonical}
  S = \int \dn{4}{x}\; \sqrt{-g}
P(\phi, \partial \phi, \partial^2 \phi , \ldots) ,  
\end{equation}
where $P$ is an arbitrary Lorentz-invariant polynomial 
of $\phi$ and its derivatives.  This form of the action includes 
the standard choice \eref{intro:scalaraction} as a special case.
Non-trivial choices of kinetic terms of the sort described by
\eref{intro:noncanonical} have been considered 
previously by a number of authors
\cite{armendariz-picon,garriga-mukhanov,hwang-noh,stewart-gong,wei-cai-wang,%
arkani-hamed-chang,arkani-hamed-creminelli}.

The presence of higher-derivative operators in $P$ allows 
for qualitatively new effects. For example, in the case where
$P$ is a function only of first derivatives of the field
(and is independent of the value of the field itself), it is possible 
for the inflaton to `condense' at a turning point in $P$.  
In this scenario, the field momentum acquires a non-zero 
vacuum expectation value, $\langle \dot{\phi} \rangle \neq 0$ 
\cite{arkani-hamed-chang,arkani-hamed-creminelli},
that is able to drive a phase of de Sitter 
(exponential) expansion. Low-energy approximations to 
stringy degrees of freedom may also 
be described through an action of the form 
\eref{intro:noncanonical}. Specific examples include 
the tachyon field or a generalized Dirac--Born--Infeld action 
\cite{garousi-sami,alishahiha-silverstein}. 
 
In view of the above possibilities, therefore, a study of more complicated 
actions of the form \eref{intro:noncanonical} is well motivated. 
In general, one may include terms with any number of higher derivatives 
in $P$.  However, in any effective theory it is 
to be expected that terms containing higher derivatives 
will be suppressed by powers of the ultraviolet cut-off scale. 
In theories coupled to Einstein gravity, this is most naturally the 
Planck scale, so if inflation occurs at energies $E \ll \planckmass$, 
the contribution from such operators eventually becomes small. This 
implies that the effect of arbitrarily higher derivatives can be neglected.  
We therefore consider the class of theories 
where $P$ contains at most first derivatives in the scalar field 
\cite{armendariz-picon,garriga-mukhanov,arkani-hamed-chang,hwang-noh,%
wei-cai-wang,stewart-gong}:
\begin{equation}
  \label{intro:pdef}
  P = P(X,\phi),  \qquad  X = -g^{ab} \grad_a \phi \grad_b \phi .
\end{equation}
Having restricted the system to first derivatives in this way, 
the requirement that these derivatives enter via $X$ is 
fixed by Lorentz invariance, so this really is the most 
general form of the Lagrangian.

Despite the novel character of \eref{intro:noncanonical}, 
and the widely differing physics that enters into the various models 
leading to such an effective Lagrangian, the predictions for 
standard observables such as the scalar spectral index are 
essentially degenerate with the standard scenario 
\eref{intro:scalaraction} 
to leading order in the slow-roll parameter $\epsilon$ 
\cite{garriga-mukhanov,wei-cai-wang}, where $\epsilon = -\dot{H}/H^2 \ll 1$.   
This implies that further observational information is required in order 
to discriminate between the alternative scenarios. One 
source of additional observational insight is provided by the three-point 
or higher connected correlation functions, 
as measured in principle through CMB fluctuations. Since the largest 
contribution is expected to arise from the three-point function, 
considerable attention has focused recently on the 
theoretical nature of this correlation and its possible   
observational detection in the CMB \cite{bartolo-matarrese-review}.

In this paper, we calculate the three-point function for a general 
theory of single-field inflation 
whose microphysics is described by the action 
\eref{intro:noncanonical}--\eref{intro:pdef}, 
using the slow-roll approximation to control the calculation 
where necessary.  
The three-point function for the canonical action 
\eref{intro:scalaraction} coupled to gravity was calculated by 
Maldacena \cite{maldacena-nongaussian} (see also \cite{acquaviva-bartolo}),
and a similar analysis has been performed by 
Rigopoulos \& Shellard \cite{rigopoulos-shellard}, 
who approximate the quantum fluctuations by a stochastic noise term.  
Some specific examples 
of theories containing higher-derivative operators have been considered 
in the literature
\cite{creminelli,arkani-hamed-creminelli,alishahiha-silverstein,gruzinov}, 
either coupled to gravity or considered in isolation.  
We perform the computation quite generally, 
including all details of the minimal coupling to Einstein gravity.
Since non-gaussianity is a potentially sensitive probe of new or unexpected
physics, some attention has also been given to non-standard scenarios,
such as tachyon or brane inflation \cite{calcagni-nongaussian}.
The case of scalar field inflation with
canonical kinetic terms coupled non-minimally to gravity was recently
considered by Koh, Kim \& Song \cite{koh-kim-song}.

One of the key features that arises in considering the 
generalized action \eref{intro:noncanonical}--\eref{intro:pdef} is that 
the speed of sound, $\cs$, is in general 
time-dependent and differs from unity (in units where 
the speed of light $c=1$), in contrast to the 
canonical action \eref{intro:scalaraction}, where $\cs =1$. 
As we shall see, this has significant implications for the 
form of the three-point correlation function. 
Our main result is that the three-point function contains terms 
that have similar $k$-dependences to  
that of standard, single-field models \cite{maldacena-nongaussian}, 
but with a different dependence on the slow-roll parameters. 
Furthermore, new $k$-dependent terms are present which are 
entirely absent in the standard case and arise whenever 
the speed of sound differs from unity. In principle, 
these new features represent a distinctive probe of 
$\cs \neq 1$ in the CMB \cite{babich-creminelli}, 
since they produce a pattern of angular dependence on 
the CMB sky which ought to be accessible whenever the 
non-gaussianity predicted by \cite{maldacena-nongaussian} 
is observable.  Our expression for the three-point function also 
respects the long-wavelength gravitational consistency 
relation \cite{maldacena-nongaussian,creminelli-zaldarriaga}.

The outline of this paper is as follows.  
We discuss the homogeneous background model in 
Section~\ref{sec:background}, establishing 
our notation for the equations of motion
and proceeding to discuss the slow-roll approximation 
for models with generalized kinetic terms. In these models,  
there is an extra requirement, over and above the familiar 
restriction that field derivatives should be less than the 
expansion rate, which follows from demanding 
that the rate of change of the sound speed should be sufficiently small.  
In Section~\ref{sec:vertex} we couple fluctuations 
in the scalar field to scalar modes of the metric.  
This is most simply expressed using the ADM decomposition \cite{salopek-bond}.  
We solve the ADM constraint equations in Section~\ref{sec:constraint}.  
Given a solution of the constraints 
it is possible to construct the Gaussian action, as shown in 
Section~\ref{sec:freefield}.  This has been done previously in the literature 
using different techniques \cite{garriga-mukhanov,hwang-noh}.  

In Section~\ref{sec:interaction}, 
we construct the interaction vertex for the coupled fluctuations.
This interaction vertex may be viewed as the 
generalization of the third-order action presented 
in \cite{maldacena-nongaussian} to include a varying speed of sound, or 
alternatively, as an extension to third-order 
of the $k$-inflation action derived by 
Garriga \& Mukhanov \cite{garriga-mukhanov}.   
We calculate the three-point function by employing a 
different technique to that most commonly used in the literature to 
date. This provides an alternative and independent method for calculating 
the three-point function which may sometimes prove more convenient.  
We outline this method in Section~\ref{sec:formalism}, and show how it 
leads to an expression for the tree-level (`semiclassical') approximation 
of the three-point function in a particularly transparent and direct
way.  The details of the calculation are presented in 
Section~\ref{sec:calc}.  We recast the result as an expression for 
the conventional non-linearity parameter $\fnl$ and show explicitly 
how the result obeys Maldacena's consistency condition
\cite{maldacena-nongaussian}.
Finally, we draw our conclusions in Section~\ref{sec:conclusions}.

\section{The background model}
\label{sec:background}
We work with an action of the form
\begin{equation}
  \label{intro:action}
  S = \frac{1}{2} \int \dn{4}{x}\; \left[ R + 2P(X,\phi) \right] , 
 \quad \quad
  X = - g^{ab} \grad_a \phi \grad_b \phi , 
\end{equation}
where units are chosen such that the reduced Planck 
mass $\planckmass^{-2} = 8\pi G$ is set to unity. 
The homogeneous background solution is assumed to be of 
Friedmann--Robertson--Walker (FRW) form with flat spatial slices,
\begin{equation}
  \label{backg:metric}
  \measure{s}^2 = - \measure{t}^2 + a^2(t) \measure{\vect{x}}^2 .
\end{equation}
Given this metric, the gravitational dynamics supply both an evolution 
equation and a constraint for the Hubble parameter, $H = \dot{a}/a$:
\begin{eqnarray}
  \label{backg:friedmann}
  2 \dot{H} + 3 H^2 = -P , \nonumber \\
  H^2 = \frac{1}{3}(2 X P_{,X} - P)  ,
\end{eqnarray}
where a comma denotes a partial derivative.  The constraint equation is 
merely the Friedmann equation in this model.  
The equation of motion for the scalar field is given by  
\begin{equation}
\label{scalarequation}
  \fl
  \dot{X} ( P_{,X} + 2X P_{,XX} ) + 2 \sqrt{3} (2XP_{,X} - P)^{1/2} X P_{,X} =
  X^{1/2} ( P_{,\phi} - 2 X P_{,X \phi})  .
\end{equation}

An important consequence
of the non-trivial kinetic structure in $P$ is that 
the na\"{\i}ve dispersion relation for $\phi$ is modified, 
and fluctuations in the scalar field do not travel at the speed of light.  
Instead, the sound speed in $\phi$ is given by 
\begin{equation}
  \label{backg:soundspeed}
  \cs^2 = \frac{P_{,X}}{P_{,X} + 2X P_{,XX}} .
\end{equation}

\subsection{The slow-roll approximation}
\label{sec:sr}
For general $P$ the scalar field equation (\ref{scalarequation}) 
cannot be solved analytically. In order to proceed, therefore, 
it is necessary to resort to approximations, where the solution
is expanded perturbatively in powers of a small parameter.
Within the context of standard scalar field inflation, 
this is usually achieved by assuming that the field $\phi$ 
is rolling slowly in comparison to the expansion rate, i.e., 
$\dot{\phi}^2 \ll H^2$.  More quantitatively, we may define
\begin{equation}
  \label{backg:epsilondef}
  \epsilon = - \frac{\dot{H}}{H^2} = \frac{X P_{,X}}{H^2}, \quad  \quad
  \eta = \frac{\dot{\epsilon}}{\epsilon H} ,
\end{equation}
with the understanding that $|\epsilon|$, $|\eta| \ll 1$ 
for reliable calculations%
\footnote{In standard,  
single-field inflation, $\epsilon$ is usually a 
positive quantity by definition.  That need not be the case here.  
There are several inequivalent definitions of $\eta$ 
which are used in the literature, of which the most common alternatives 
to our choice are $\eta_V = V''/V$ \cite{liddle-lyth} 
and $\eta_H = 2 H''/H$ \cite{lidsey-liddle}.  The former 
definition makes sense only for standard inflation, whereas 
the latter can be used where non-trivial kinetic terms are present.  
With a canonical choice of kinetic term, one can show that these 
alternatives are related to our $\eta$ by the rules
\begin{equation}
  \eta = - 2 \eta_H + 2 \epsilon = -2 \eta_V + 4 \epsilon .
\end{equation}
Note that these \emph{only} apply for standard inflation.}.
In practice we will assume that $\epsilon \sim \eta$ and 
express this condition by writing 
$\epsilon, \eta \sim \Or(\epsilon)$.  

It proves useful to 
decompose the parameter $\epsilon$ into two new dimensionless ratios, 
$\epsilon_\phi$ and $\epsilon_X$, which measure how the expansion rate 
varies with the kinetic and potential parts of $\phi$, respectively: 
\begin{equation}
  \epsilon = -\frac{\dot{\phi}}{H^2}\pd{H}{\phi} - 
\frac{\dot{X}}{H^2}\pd{H}{X} = \epsilon_\phi + \epsilon_X .
\end{equation}
The scalar field equation of motion (\ref{scalarequation}) may then 
be written as 
\begin{equation}
  \label{backg:fieldeqn}
  \dot{X} = - 6 H \cs^2 X \left(1 - \frac{\epsilon_\phi}{\epsilon} \right) ,
\end{equation}
and this allows us to express $t$-derivatives in terms of derivatives with 
respect to $X$. 

In principle, there is no requirement from a dynamical 
point of view that $\epsilon_\phi$ and $\epsilon_X$ 
should both be small, even when $|\epsilon| \ll 1$.  
In standard, single-field inflation, $\epsilon$ and $\eta$ are often 
referred to as the \emph{slow-roll} parameters, and the limit 
$|\epsilon|$, $|\eta| \ll 1$ as the \emph{slow-roll limit}.  
This terminology is not quite appropriate for a general choice of $P$, 
since $|\epsilon| \ll 1$ no longer entails $\dot{\phi}^2 \ll H^2$.  
For brevity, however, we can refer to $\epsilon$ and $\eta$ 
as \emph{flow} parameters, since they describe how the theory evolves 
on the space of inflationary models \cite{hoffman-turner,liddle-flow}.  
By an abuse of terminology, we will continue to describe the 
limit $|\epsilon| \ll 1$ as `slow-roll' because the content of the 
approximation is familiar in the literature.

As well as the familiar conditions 
$|\epsilon|$, $|\eta| \ll 1$, it will also be necessary impose 
bounds on the rate of change of the sound speed due to the generalized 
kinetic terms in \eref{intro:action} 
\cite{garriga-mukhanov,hwang-noh,wei-cai-wang,stewart-gong}.  
We therefore define the parameters
\begin{equation}
  u = 1 - \frac{1}{\cs^2} = -2 X \frac{P_{,XX}}{P_{,X}} , \quad \quad
  s = \frac{1}{H} \frac{\dot{\cs}}{\cs} ,
\end{equation}
where $s$ represents a dimensionless measure of the rate of change
of the sound speed $\cs^2 =1/(1-u)$. 
These two parameters are related by 
\begin{equation}
  \label{backg:udot}
  \dot{u} = 2Hs(1-u) .
\end{equation}
It is well-known that the time derivatives of $\epsilon$ and 
$\eta$ are second-order in the slow-roll expansion, in the sense 
that $\dot{\epsilon}$, $\dot{\eta} \sim \Or(\epsilon^2)$ 
\cite{lidsey-liddle}.  This means that we can consistently work to 
first-order in $\Or(\epsilon)$, while keeping 
$\epsilon$ and $\eta$ constant.  
Eq.~\eref{backg:udot} implies that $s$ is related to the 
time derivative of $u$, so it is sufficient that
$u = \Or(\epsilon)$ in order that 
$s = \Or(\epsilon^2)$. In this case, 
$\cs^2$ departs from unity only by a quantity that is 
first-order in slow-roll.

After combining \eref{backg:epsilondef} with 
the scalar field equation \eref{backg:fieldeqn},  
we may write down a relationship between the parameters 
$\epsilon$, $\eta$ and $u$: 
\begin{equation}
  u = \frac{\epsilon(2\epsilon - \eta) - 6 \epsilon_X}
  {\epsilon(2\epsilon - \eta) - 3 \epsilon_X} .
\end{equation}
It follows that a necessary condition for $u = \Or(\epsilon)$
is that $\epsilon_X$ satisfies
\begin{equation}
  \label{intro:epsilonx}
  6 \epsilon_X = \epsilon(2\epsilon - \eta) + \Or(\epsilon^3) 
\end{equation}
and this implies that $\epsilon_X$ is subdominant with respect to 
$\epsilon_\phi$. 
Indirectly, this is a rather non-trivial condition on $P(X,\phi)$ 
and means that no guarantee can be given that a particular $P(X,\phi)$ 
will necessarily support a phase where $|\epsilon| \ll 1$, 
even in principle.  In this paper, we do not attempt to ascertain 
the conditions under which a particular $P(X,\phi)$ will admit a 
slow-roll epoch, but merely provide an expression for the three-point 
function which is valid whenever it does.

In order to simplify some of the formulae that follow, 
it will prove useful to introduce two new quantities, 
$\Sigma$ and $\lambda$, which are combinations of derivatives of 
$P$, and defined by
\begin{equation}
  \label{backg:sigmalambda}
  \Sigma = X P_{,X} + 2 X^2 P_{,XX}, \quad   \quad
  \lambda = X^2 P_{,XX} + \frac{2}{3} X^3 P_{,XXX} .
\end{equation}
These can be written in terms of flow parameters:
\begin{equation}
  \label{backg:sigmalambdasr}
  \Sigma = \frac{H^2 \epsilon}{\cs^2} = H^2 \epsilon (1-u) , \quad \quad 
  \lambda = \frac{\Sigma}{6} \left[ \frac{2}{3}\frac{\epsilon}{\epsilon_X}
  (1-u) s - u \right] .
\end{equation}

\section{The ADM formalism}
\label{sec:vertex}
Any consistent cosmological calculation of fluctuations in some scalar 
field $\phi$ which dominates the energy density of the universe must 
account for the universal coupling to gravity, since any perturbation in 
$\phi$ will produce a non-negligible perturbation in the energy--momentum 
tensor.  Thus, we need to calculate the action for small 
fluctuations around the homogeneous background solution 
of \eref{intro:action}, taking into account both the perturbations 
in the scalar field, $\delta\phi$, and the scalar modes of the metric.  
There is no need to include vector perturbations, which die away rapidly 
with the cosmic expansion and are not sourced by inflation.  
In addition, we omit tensor modes.  In principle, tensor modes 
corresponding to gravitational waves are excited by inflation, 
but gravitational waves have not yet been detected and it is 
anticipated that any non-gaussianity involving such modes 
will be at a lower level than that predicted for the scalar sector
\cite{maldacena-nongaussian}.  
In the near future, observational effort is likely to be directed towards 
the determination of the scalar non-gaussianity, 
to which we restrict our attention.

An arbitrary scalar perturbation of the background \eref{backg:metric} 
can be written in the form 
\begin{equation}
  \label{vertex:scalar-perturb}
  \fl
  \measure{s}^2 = -(1+2\Phi) \measure{t}^2 + 2a^2(t) B_{,i} \, \measure{x}^i
  \measure{t} + a^2(t) \left[
  (1-2\Psi) \delta_{ij} + 2 E_{,ij} \right] \measure{x}^i\measure{x}^j ,
\end{equation}
where a comma denotes a partial derivative with respect to the 
spatial coordinates $x^i$.  One could directly calculate the action 
for the fields $\Phi$, $B$, $\Psi$ and $E$ and work with these 
fluctuations together with fluctuations $\delta\phi$ in the inflaton.  
After integrating by parts, dropping total derivatives, 
applying the constraint equations and using the background equations of 
motion, it can be shown that to quadratic order 
the action for these fluctuations can be written 
in terms of the comoving curvature perturbation $\R$ 
\cite{mukhanov-a,mukhanov-b,sasaki,mfb}:
\begin{equation}
  \label{vertex:comoving-curvature}
  \R = - \Psi - \frac{H}{\dot{\phi}} \delta\phi ,
\end{equation}
which is gauge-invariant under reparameterizations of time.  In practice, 
and especially when carrying the calculation to third order, 
it is much simpler to work in the ADM formalism \cite{salopek-bond}, where the
metric  has the form
\begin{equation}
  \label{vertex:adm}
  \measure{s}^2 = - N^2 \measure{t}^2 + h_{ij} (\measure{x}^i + N^i
  \measure{t})(\measure{x}^j + N^j \measure{t}) .
\end{equation}
In this representation $h_{ij}$ is the three-dimensional metric on slices 
of constant $t$.  The lapse function $N$ and shift vector $N^i$ 
contain the same information as the metric fields $\Phi$ and $B$.  
However, they are chosen in such a way that they appear as 
Lagrange multipliers in the action, so their equations of motion are 
purely algebraic.  After solving these constraint equations, 
the solutions for $N$ and $N^i$ can be substituted 
back into the action, thereby avoiding 
the very lengthy manipulations involved 
when working with 
\eref{vertex:scalar-perturb}--\eref{vertex:comoving-curvature}.

All our calculations simplify considerably by working in the 
comoving gauge, where the three-dimensional slices implicit 
in \eref{vertex:adm} are chosen so that the inflaton perturbation 
$\delta\phi$ vanishes.  On slices where $\delta \phi =0$, 
the three-dimensional metric takes the 
form \cite{maldacena-nongaussian,lyth-rodriguez}%
\footnote{Our notation is chosen to correspond to \cite{wands-malik}, 
where the symbol $\R$ is used for the curvature perturbation in 
the comoving gauge and $\zeta$ is used for the 
curvature perturbation in the uniform density gauge.  
In standard inflation these coincide up to choices for signs, 
but this need not be the case once a  non-trivial kinetic structure 
has been introduced into the Lagrangian.}
\begin{equation}
  h_{ij} = a^2(t) \e{2\R} \delta_{ij} ,
\end{equation}
where the field $E$ has been gauged away by 
an appropriate choice of the coordinates $x^i$, and $\R$ is the 
non-linear generalization of the comoving curvature perturbation
\eref{vertex:comoving-curvature}.%
\footnote{The reader is warned that different conventions 
for extending \eref{vertex:comoving-curvature} beyond linear order are 
employed in the literature.  The situation is nicely reviewed in
\cite{lyth-rodriguez}.}
Although in principle this is a gauge choice, our 
results will be gauge invariant up to reparameterizations 
of the spatial coordinates.  
This follows since the quantity $\R$ is actually gauge-invariant 
to all orders, being defined by the physical condition that comoving observers
see vanishing momentum flux \cite{liddle-lyth-paper,wands-malik}.  We apply the
comoving gauge uniformly throughout the present paper.  In principle there is
some interest attached to working with other gauges, such as the spatially flat
gauge or uniform density gauge, but in such cases the formalism we will
describe becomes burdened with a large number of extra terms.  These terms
arise from spatial derivatives associated with inhomogeneities which are
generically present in $\phi$, but are absent in the comoving gauge where
$\delta\phi = 0$.

\subsection{The constraint equations}
\label{sec:constraint}
With the ADM metric \eref{vertex:adm}, the coupled action 
\eref{intro:action} reduces to
\begin{equation}
  \label{vertex:action}
  \fl
  S = \frac{1}{2} \int \measure{t} \, \dn{3}{x} \; \sqrt{h} N \left( R^{(3)} +
  2 P \right) +
  \frac{1}{2} \int \measure{t} \, \dn{3}{x} \; \sqrt{h} N^{-1} (E_{ij} E^{ij}
  - E^2) ,
\end{equation}
where $h = \det h_{ij}$ and $R^{(3)}$ is the Ricci curvature calculated with
$h_{ij}$.  The symmetric tensor $E_{ij}$ is proportional to the extrinsic
curvature of the spatial slices,
\begin{equation}
  E_{ij} = \frac{1}{2} \dot{h}_{ij} - N_{(i|j)}
\end{equation}
where $|$ is the covariant derivative compatible with $h_{ij}$.  
The $N$ and $N^i$ constraint equations are
\begin{equation}
  \label{vertex:lapseconstraint}
  R^{(3)} + 2P - 4X P_{,X} - \frac{1}{N^2} (E_{ij} E^{ij} - E^2) = 0 , 
\end{equation}
\begin{equation}
  \label{vertex:shiftconstraint}
  \left[ \frac{1}{N} (E_i^j - E \delta^j_i) \right]_{|j} = 0 ,
\end{equation}
respectively.

In solving these equations, 
we follow \cite{maldacena-nongaussian} and split the shift vector $N_i$ 
into irrotational and incompressible parts, $N_i = \psi_{,i} + \tilde{N}_i$,
where $\tilde{N}_{i,i} = 0$.  After setting $N = 1 + \alpha$, the quantities
$\alpha$, $\psi$ and $\tilde{N}_i$ admit expansions into powers of $\R$,
\begin{eqnarray}
\nonumber
\alpha = \alpha_1 + \alpha_2 + \cdots, \\
\nonumber \psi = \psi_1 + \psi_2 + \cdots,  \\ 
\tilde{N}_i = \tilde{N}_i^{(1)} + \tilde{N}_i^{(2)} + \cdots ,
\end{eqnarray}
where (for example) $\alpha_n = \Or(\R^n)$. We
then set the constraints to zero order-by-order.  
The background equation is the Friedmann equation \eref{backg:friedmann}.  
At first-order, one finds from the $N^i$ constraint that
\begin{equation}
  \label{vertex:constrainta}
  \alpha_1 = \frac{\dot{\R}}{H} , \qquad
  \partial^2 N_i^{(1)} = 0 ,
\end{equation}
so with an appropriate choice of boundary conditions one can justifiably 
set $N_i^{(1)} = 0$. It follows from the $N$ constraint that%
\footnote{The operator $\partial^{-2}$ is the solution operator for the
Laplacian, defined by $\partial^{-2}(\partial^2 \phi) = \phi$.}
\begin{equation}
  \label{vertex:constraintb}
  \psi_1 = - \frac{\R}{H} + \frac{a^2}{H^2}\Sigma \; \partial^{-2} \dot{\R} ,
\end{equation}
where $\Sigma$ was defined in \eref{backg:sigmalambda}.  
As emphasized in \cite{maldacena-nongaussian}, when calculating the 
action to order $n$ in $\R$, we do not need to compute 
the order-$\R^n$ term in $N$ or $N^i$, since this must be multiplying 
$\partial L/\partial N$ or $\partial L/\partial N^i$ 
and these are both zero by virtue of the constraint equations.  
In general, one would need all terms up to and including 
$\Or(\R^{n-1})$, but in the present case, 
terms of order $\R^2$ drop out of the third-order action, 
so \eref{vertex:constrainta}--\eref{vertex:constraintb} are sufficient 
to calculate the order-$\R^3$ term.

\subsection{The free field action}
\label{sec:freefield}
Using \eref{vertex:constrainta}--\eref{vertex:constraintb} to 
solve for $N$ and $N^i$ in the action and keeping terms up to 
quadratic order in $\R$, the second-order action is
\begin{equation}
  \label{vertex:freeaction}
  S_2 = \int \measure{\tau}\,\dn{3}{x}\; a^2 \left[ \frac{\Sigma}{H^2}
  (\R')^2 - \epsilon (\partial \R)^2 \right] ,
\end{equation}
in agreement with the action for $k$-inflation calculated by 
Garriga \& Mukhanov \cite{garriga-mukhanov}, 
where $\tau$ denotes conformal time,   
defined by $\measure{t} = a \, \measure{\tau}$, and 
a prime $'$ denotes a derivative with respect to $\tau$.  
Conformal time during inflation is given to leading order in slow-roll
by $\tau = -(aH)^{-1}$.

In practice, it is convenient to introduce a rescaled field 
$v = z \R$, where $z$ is defined by 
\begin{equation}
  z^2 = \frac{2a^2 \Sigma}{H^2} = \frac{2a^2\epsilon}{\cs^2}
\end{equation}
and the speed of sound $\cs$ was defined in \eref{backg:soundspeed}.  
In terms of $v$ the action becomes $\frac{1}{2}\int v \triangle v$, 
where the operator $\triangle$ satisfies
\begin{equation}
  \triangle = -\frac{\partial^2}{\partial \tau^2} + \cs^2 \partial^2 + 
  \frac{z''}{z} .
\end{equation}
The $v$-propagator between time $\tau_0$ and time $\tau$ is 
$G_v(\tau,\tau_0) = \imag \triangle^{-1}(\tau,\tau_0)$, 
where to reduce clutter in the notation we have suppressed the 
spatial dependence in $\triangle$.  When expressed in terms of Fourier 
modes, this simply means that 
\begin{equation}
    G_v(\tau,\tau_0) = \int \frac{\dn{3}{k}}{(2\pi)^3} \; 
    G_v(k,\tau)\e{-\imag \vect{k}\cdot(\vect{x}-\vect{y})}, 
\end{equation}
where 
\begin{equation}
\label{muk}
  G_v'' + \left( \cs^2 k^2 - \frac{z''}{z} \right) G_v = -\delta(\tau-\tau_0) .
\end{equation}
Eq. \eref{muk} is known as the Mukhanov 
equation and is equivalent to $\triangle v =0$
\cite{mukhanov-a,mukhanov-b,mfb}. 

In general, this equation for $G_v$ is not easy to solve.  
The effective mass $z''/z$ can be expressed in the form
\begin{equation}
  \frac{z''}{z} = \frac{3/2 + \nu}{\tau^2} 
\end{equation}
where $\nu$ is a combination of 
terms that are linear and quadratic in the
slow-roll parameters \cite{garriga-mukhanov,lidsey-liddle}. 
To obtain an approximate solution in standard inflation, that is
valid to first-order in slow-roll, terms in $\nu$ that are 
first-order in $\epsilon$ are treated as constants
and terms of $\Or(\epsilon^2)$ are dropped.  
This procedure only makes sense if
the time derivatives of the $\Or (\epsilon )$ 
quantities may be neglected along with the $\Or(\epsilon^2)$ terms, 
which requires $|\epsilon|$, $|\eta| \ll 1$.  
Of course, this is nothing more than the familiar 
slow-roll approximation of standard inflation. 
However, in the present context there is an extra condition 
arising from the requirement that $\cs$ must also 
be kept constant%
\footnote{The much more complicated case where $\cs$ may 
have some appreciable evolution was considered in 
\cite{wei-cai-wang,stewart-gong}.}. 
The error arising from this latter approximation will be at least as 
significant as that arising from the mass term, 
so consistency requires 
that $s = \Or(\epsilon^2)$. After  
taking into account Eq. \eref{backg:udot}, this implies 
that $u$ must itself be of order $\epsilon$. 
In other words, the approximate solutions 
of \eref{muk} are only valid if $\cs$ is 
sufficiently close to unity, to 
within a quantity that is small in the slow-roll limit. 

To first-order in this generalized sense of slow-roll, 
the $\R$-propagator satisfies  
\begin{equation}
  \label{vertex:wavefn}
  \fl
  G_{\R}(\tau,\tau_0) = \frac{H^2}{4\epsilon\cs}\frac{1}{k^3} \times \left\{
  \begin{array}{l@{\hspace{5mm}}l}
  \displaystyle
  (1-\imag k \cs \tau_0)(1+\imag k \cs \tau)\e{-\imag k \cs (\tau-\tau_0)}
  & \tau > \tau_0 \\
  \displaystyle
  (1+\imag k \cs \tau_0)(1-\imag k \cs \tau)\e{\imag k \cs(\tau-\tau_0)}
  & \tau < \tau_0
  \end{array}\right. ,
\end{equation}
where we have chosen boundary conditions so that $G_{\R}$ 
behaves like the flat space propagator at very early times, 
when the mode is deep within the horizon and cannot 
feel the curvature of spacetime.  
This corresponds to the Bunch--Davies vacuum \cite{birrell-davies}.

The power spectrum of $\R$ is easily obtained from \eref{vertex:wavefn} 
and was derived in \cite{garriga-mukhanov}.  On large scales, dropping less
singular pieces as $k \rightarrow 0$, the two-point function is%
\footnote{Our expression for the power spectrum differs by an overall 
factor of $\cs^{-1}$ from the corresponding expression 
in \cite{garriga-mukhanov}.  At leading order in slow-roll this is 
harmless, because we have already seen that $\cs$ must be equal 
to unity to within $\Or(\epsilon)$.  
However, if one wishes to expand consistently in the small parameters 
of the approximation, one should set $\cs = 1$ exactly 
to leading order in \eref{vertex:twopt}. (See \cite{wei-cai-wang} for a 
more detailed discussion of this point.)}
\begin{equation}
  \label{vertex:twopt}
  \langle \R(\vect{k}_1) \R(\vect{k}_2) \rangle = (2\pi)^3 
\delta(\vect{k}_1 + \vect{k}_2) P(\vect{k}_1), \quad
  \quad
  P(\vect{k}_1) = \frac{H^2}{4 \epsilon} \frac{1}{k_1^3} ,
\end{equation}
where $k = |\vect{k}|$.  
To turn this into a power spectrum one takes the coincidence 
limit to find the dispersion, $\sigma^2 = \langle \R(x)^2 \rangle$,
and then evaluates its logarithmic derivative: 
\begin{equation}
  \Delta^2(k) = \deriv{\sigma^2}{\,\ln k} = 
  \frac{1}{8\pi^2} \frac{H^2}{\epsilon} .
\end{equation}
The tilt of this spectrum is given by
\begin{equation}
  \label{vertex:tilt}
  n - 1 = \deriv{\,\ln \Delta^2(k)}{\,\ln k} \approx -2 \epsilon - \eta .
\end{equation}

The quantity $\R$ is conserved outside the horizon 
\cite{garriga-mukhanov,vernizzi}, which is the analogous result 
to that of conventional slow-roll inflation \cite{wands-malik,lyth-malik}.

\section{The third-order action}
\label{sec:interaction}

\subsection{General form of the action}

We now turn to the central calculation of the present paper, 
a determination of the third order piece in the coupled action 
\eref{vertex:action}.  In order to compute this, 
we need expressions for $R^{(3)}$, $P(X)$ and $E^{ij}E_{ij} - E^2$.  
It is easy to show that 
$R^{(3)} = - 2a^{-2}\e{-2\R}[ (\partial \R)^2 + 2 \partial^2 \R ]$.  
To calculate $P(X,\phi)$, we use the fact that
\begin{equation}
  \frac{1}{N^2} = 1 - 2 \frac{\dot{\R}}{H} - 2 \alpha_2 +
  3 \frac{\dot{\R}^2}{H^2} + 6 \alpha_2 \frac{\dot{\R}}{H} -
  4 \frac{\dot{\R}^3}{H^3} .
\end{equation}
Since we have chosen a gauge in which $\delta\phi = 0$, this means
\begin{eqnarray}
  \nonumber
  \fl
  P(X,\phi) = P + X \left( -2 \frac{\dot{\R}}{H} - 2 \alpha_2 +
  3 \frac{\dot{\R}^2}{H^2} + 6 \alpha_2 \frac{\dot{\R}}{H} -
  4 \frac{\dot{\R}^3}{H^3} \right) P_{,X} \\
  + \frac{1}{2} X^2\left( 4 \frac{\dot{\R}^2}{H^2} +
  8 \alpha_2 \frac{\dot{\R}}{H} -
  12 \frac{\dot{\R}^3}{H^3} \right) P_{,XX} - \frac{8}{3!} X^3 P_{,XXX}
  \frac{\dot{\R}^3}{H^3} .
\end{eqnarray}
No derivatives of $P$ with respect to $\phi$ occur, since the inflaton 
field takes its unperturbed value.  This is one convenience of 
working in the gauge $\delta\phi=0$.  Finally, using the connection 
derived from $h_{ij}$, and our choices for $N$ and $N_i$, we obtain
\begin{eqnarray}
  \nonumber
  \fl
  E^{ij}E_{ij} - E^2 = - 6(H+\dot{\R})^2 + \frac{4H}{a^2}
  (1+\frac{\dot{\R}}{H})\e{-2\R}(\partial^2 \psi +
  \partial\R\cdot\partial\psi + \tilde{N}\cdot\partial\R) \\ \nonumber
  - \frac{1}{a^4}\e{-4\R}(\partial^2\psi\partial^2\psi +
  2 \partial^2\psi [\partial\R\cdot\partial\psi]) \\
  + \frac{1}{a^4}\e{-4\R}(\psi_{,ij}\psi_{,ij} + 2 \psi_{,ij}\tilde{N}_{i,j} -
  4 \R_{,i} \psi_{,ij}\psi_{,j} +
  2\partial^2\psi [ \partial\R\cdot\partial\psi ] ) .
\end{eqnarray}

Once one has collected terms and integrated by parts where possible, 
it turns out that all terms involving $\psi_2$ and $\tilde{N}_i^{(2)}$ 
either cancel among themselves or reduce to total derivatives, which 
can be discarded.  The remaining second-order 
contribution from the ADM quantities $N$ or $N^i$ is just 
proportional to $\alpha_2$, and can be written as
\begin{equation}
  S_{\mathrm{ADM}^2} = - \frac{1}{2} \int \measure{\tau}\,\dn{3}{x}\;
  4Ha \left( \partial^2 \psi_1 -
  \frac{1}{H} \partial^2 \R + \frac{a^2}{H^2} \Sigma \dot{\R} \right) .
\end{equation}
This vanishes when $\psi_1$ takes its on-shell 
value \eref{vertex:constraintb}, since it is proportional 
to a constraint.  (This justifies the statement made in 
Section~\ref{sec:constraint} that it is only necessary to calculate 
$N$ and $N_i$ to first-order.)  
It is most economical to rewrite the resulting action in terms of the 
quantities $\Sigma$ and $\lambda$ of \eref{backg:sigmalambda}.  
After integrating by parts, discarding total derivatives, 
and using the background equations of motion, the third-order contribution 
to the action can be written in the form 
\begin{eqnarray}
  \nonumber
  \fl
  S_3 = \frac{1}{2} \int \measure{\tau}\,\dn{3}{x}\; a^3
  \Bigg[ \frac{2}{a^2} \frac{\dot{H}}{H^2} \R(\partial\R)^2 -
  (2\Sigma + 4 \lambda) \frac{\dot{\R}^3}{H^3} +
  6 \Sigma \frac{\R\dot{\R}^2}{H^2}  -
  \frac{4}{a^4}\partial^2\psi_1 \R_{,i}\psi_{1,i} \\
  \label{vertex:actiona}
  \mbox{} - \frac{3}{a^4} \R \partial^2\psi_1\partial^2\psi_1 +
  \frac{1}{a^4}\frac{\dot{\R}}{H} +
  \frac{3}{a^4}\R \psi_{1,ij}\psi_{1,ij} -
  \frac{1}{a^4}\frac{\dot{\R}}{H}\psi_{1,ij}\psi_{1,ij} \Bigg]  .
\end{eqnarray}

To complete the reduction, we must replace 
$\psi_1$ with its value, given by Eq. 
\eref{vertex:constraintb}.  In doing so, 
it is very convenient to make use of the equation of 
motion derived from the free, Gaussian theory defined 
by \eref{vertex:freeaction}.  If we introduce a new quantity $\Lambda$, 
satisfying
\begin{equation}
  \Lambda = \frac{a^2}{H^2} \Sigma \dot{\R} ,
\end{equation}
the field equation which follows from \eref{vertex:freeaction} 
can be written as
\begin{equation}
  \label{vertex:eom}
  \left.\frac{\delta L}{\delta \R}\right|_1 =
  \deriv{\Lambda}{t} + H \Lambda - \epsilon \partial^2\R .
\end{equation}
This vanishes when $\R$ is a field mode which solves the 
equation of motion \eref{muk} of the Gaussian theory, 
but $\delta L/\delta \R|_1$ will be non-zero when $\R$ 
satisfies the equation of motion of the full 
interacting theory that takes into account the $\R^3$ vertex that we are 
computing. By using \eref{vertex:eom}, we can trade derivatives of 
$\Lambda$, which will appear after integrating by parts in 
\eref{vertex:actiona}, for simpler terms involving $H\Lambda$, 
$\partial^2 \R$ and $\delta L/\delta \R|_1$.  Eventually, we will be able 
to remove the $\delta L/\delta \R|_1$ terms by a field redefinition, 
resulting in a considerable simplification of the $\R^3$ vertex.

Proceeding in this way, one finds that the action can be written in the form 
\begin{eqnarray}
  \nonumber
  \fl
  S_3 = \frac{1}{2} \int \measure{\tau}\,\dn{3}{x}\; a^3\Bigg( - \frac{2}{a^2}
  \epsilon \R(\partial\R)^2 -
  (2\Sigma + 4 \lambda) \frac{\dot{\R}^3}{H^3} + 6 \Sigma \R \dot{\R}^2
  \\ \nonumber 
  \mbox{} + \frac{2}{a^4} \frac{\Lambda}{H}(\partial \R)^2 -
  \frac{4}{a^4} \Lambda \R_{,i} \chi_{,i} + \frac{1}{a^4} \Lambda \epsilon
  \R_{,i} \chi_{,i} + \frac{\epsilon}{2a}
  \partial^2 \R (\partial \chi)^2 \\ \label{vertex:auxfinal}
  \mbox{} + \mbox{terms involving $\delta L/\delta \R|_1$} \Bigg)  ,
\end{eqnarray}
where the terms involving $\delta L/\delta \R|_1$ are given by
\begin{eqnarray}
  \label{vertex:auxredef}
  \nonumber \fl
  \frac{1}{a^4 H^2} \Bigg[ ( \R_{,i} \chi_{,i} - [\partial \R]^2 )
  \left.\frac{\delta L}{\delta\R}\right|_1
  + (\partial^2 \R \chi_{,i} + \Lambda \R_{,i} ) \left( \partial^{-2}
  \left.\frac{\delta L}{\delta\R}\right|_1 \right)_{,i} 
  \\
  \mbox{} - \frac{1}{H} \R_{,i} \R_{,j}
  \left( \partial^{-2}\left.\frac{\delta L}{\delta\R}\right|_1\right)_{,ij}
  \Bigg]
\end{eqnarray}
and we have set $\chi = \partial^{-2} \Lambda$.  These terms will not be 
important in what follows, because they all contain at least one 
derivative of $\R$ and therefore vanish outside the horizon, where $\R$
approaches 
a constant.  When determining the order in slow roll of each of the terms 
in \eref{vertex:auxfinal}--\eref{vertex:auxredef}, it should be 
noted that $\chi = \Or(\epsilon)$: $\chi$ (and $\Lambda$) 
are first-order in slow-roll.  

After further integrations by parts and 
use of the Gaussian field equation \eref{vertex:eom}, we find that the 
entire three-point vertex can be expressed in terms of the flow 
parameters $\epsilon$, $\eta$, $\epsilon_X$, $u$ and $s$:
\begin{eqnarray}
  \label{vertex:final}
  \nonumber
  \fl
  S_3 = \frac{1}{2} \int \measure{t}\,\dn{3}{x}\; a^3
  \Bigg[ -\frac{4}{3}\epsilon\left( \frac{1}{3\cs^2}
  \frac{\epsilon}{\epsilon_X} s + u \right)
  \frac{\dot{\R}^3}{H \cs^2} + \frac{2\epsilon}{\cs^2}
  \left( 3 u + \frac{\epsilon}{\cs^2} \right) \R \dot{\R}^2
  \\ \nonumber +
  \frac{2\epsilon}{a^2 \cs^2} ( \epsilon - 2 s - u \cs^2 )
  \R (\partial \R)^2 \\
  - \frac{4}{a^2} \frac{\epsilon}{\cs^2} \dot{\R}\R_{,i}\chi_{,i} +
  \R^2 \dot{\R}
  \frac{\epsilon}{\cs^2} \deriv{}{t} \left( \frac{\eta}{\cs^2} \right) -
  \frac{\epsilon^3}{\cs^4} \R \dot{\R}^2 + \frac{1}{a^2} \epsilon \R
  \chi_{,ij} \chi_{,ij} \Bigg] .
\end{eqnarray}
Although this has the appearance of a series expansion in terms of 
flow parameters, this expression 
is in fact \emph{exact} to $\Or(\R^3)$, 
given that interactions of both gravity and the scalar field  
with the other contents of the universe have been neglected.  
This vertex should be supplemented by terms proportional to 
$\delta L/\delta \R|_1$:
\begin{equation}
  \label{vertex:redef}
  S_{3,\mathrm{Gaussian}} = \frac{1}{2}\int \measure{t}\,\dn{3}{x}\;
  \frac{\eta a}{\cs^2} \R^2 \left.
  \frac{\delta L}{\delta \R}\right|_1 + \cdots ,
\end{equation}
where we have omitted the derivative terms in \eref{vertex:auxredef}, 
which vanish outside the horizon.

Each of the parameters appearing in \eref{vertex:final} 
have comparatively simple interpretations.  The parameters $\epsilon$ and 
$\eta$ describe the non-gaussianity produced by coupling a scalar field 
(with \emph{any} given kinetic structure or self-interaction
\cite{maldacena-nongaussian}) to 
gravity.  Terms containing $u$ measure how far the sound speed deviates 
from the speed of light: in other words, a dispersion relation different 
from the canonical case $E^2 = p^2 + m^2$ acts as a source for 
non-gaussianity.  There is another source of non-gaussianity which 
arises from \emph{changes} in the speed of sound, as measured by $s$.  
This particular source appears in combination with $\epsilon/\epsilon_X$, 
which measures the non-trivial nature 
of the kinetic structure in $P(X,\phi)$.

\subsection{Slow-roll limit}

We only calculate to leading order in slow-roll.  
From a practical point of view, this is necessary given the complexity of the 
calculation beyond 
leading order, but since the predicted non-gaussianity is rather small 
\cite{maldacena-nongaussian}, only the leading-order effect 
is ever likely to be observed.  
Restricting \eref{vertex:final} to leading order terms -- 
in this case, they are $\Or(\epsilon^2)$ -- 
we find that the three-point vertex is well-approximated 
in the slow-roll r\'{e}gime by the expression 
\begin{eqnarray}
  \label{threept:vertex}
  \fl
  S_3 = \frac{1}{2}\int \measure{t}\,\dn{3}{x}\; a^3 \left[
  -\frac{4}{3}\epsilon \left( u + \frac{\epsilon}{\epsilon_X}\frac{s}{3} 
  \right) \frac{\dot{\R}^3}{H}
  + 2 \epsilon (3u + \epsilon) \R\dot{\R}^2 \right. \nonumber \\
  \left. 
  + \frac{2}{a^2}\epsilon(\epsilon - u) \R (\partial\R)^2
  - \frac{4}{a^2} \epsilon \dot{\R} \R_{,i} \chi_{,i} \right] ,
\end{eqnarray}
together with one supplementary term proportional to 
$\delta L/\delta \R|_1$, as given by Eq. \eref{vertex:redef}. 
This is also a leading-order term in slow roll.  
Despite appearances, the term involving $s$ really 
is of order $\epsilon^2$, 
since we have already seen from Eq. \eref{intro:epsilonx} 
that $\epsilon_X = \Or(\epsilon^2)$ 
whenever a slow-roll r\'{e}gime exists.  
Although in principle we could proceed to the calculation 
with this vertex in its present form, it is worth performing a 
further integration by parts to remove the term 
involving $\chi_{,i}$.  After doing this,
and using the Gaussian field equation \eref{vertex:eom}, 
we can rewrite \eref{threept:vertex} as
\begin{eqnarray}
  \label{threept:reduced}
  \nonumber \fl
  S_3 = \int \measure{t}\,\dn{3}{x}\; \Bigg[ -\frac{2}{3}a^3 \epsilon
  \left( u + \frac{\epsilon}{\epsilon_X}\frac{s}{3}
  \right) \frac{\dot{\R}^3}{H} +
  4 a^5 \epsilon(2u + \epsilon)H \dot{\R}^2\partial^{-2}\dot{\R} \\
  \mbox{} - 4a^3
  \epsilon\dot{\R}\partial^2\R\partial^{-2}\dot{\R} \Bigg] ,
\end{eqnarray}
together with some terms that are proportional to the 
Gaussian equations of motion,
\begin{equation}
  \label{threept:redef}
  S_{3,\mathrm{Gaussian}} = \frac{1}{2} \int \measure{t}\, \dn{3}{x}\; a
  \mathcal{F} \left. \frac{\delta L}{\delta \R}\right|_1, 
\end{equation}
where 
\begin{equation}
\label{defF}
\mathcal{F} = (\eta - u - \epsilon)\R^2 + 2\epsilon 
\partial^{-2}(\R \partial^2 \R) 
\end{equation}
includes the contribution of \eref{vertex:redef}. 
There are also terms that are of order 
$\Or(\epsilon^3)$ and higher, as well as 
terms which are proportional to the equations 
of motion, but these are irrelevant since they contain derivatives of $\R$.

Finally, we should confirm that the form of this vertex 
reproduces known results in standard, single field inflation.  
In fact, 
assuming the canonical form $P(X,\phi) = \frac{1}{2}X - V(\phi)$ 
for the polynomial $P$,
it is easy to show that \eref{threept:reduced}--\eref{defF}, 
reproduce the corresponding results of \cite{maldacena-nongaussian}.

\subsection{Field redefinitions}
\label{sec:redef}

Before describing the calculation in detail, we first show that a field 
redefinition of the form $\R \mapsto \R_n + F(\R_n)$ can be used to 
eliminate terms proportional to the Gaussian equation of motion, 
$\delta L/\delta \R|_1$.  In principle we could evaluate these terms with the
others, but the most economical way of accounting for them is to 
remove all such terms via a suitable field redefinition, and then 
incorporate their effect into the correlation function by applying 
Wick's theorem.  Any terms which vanish outside the horizon, 
such as those omitted in \eref{vertex:redef} or \eref{threept:redef}, will make 
no contribution to the correlation functions on superhorizon scales.  
Hence, they may be legitimately ignored.  

A field redefinition of the form 
$\R \mapsto \R_n + F(\R_n)$, where $F$ is quadratic in $\R_n$, 
has no effect on any of 
the $\Or(\R^3)$ terms in \eref{threept:reduced}.  
On the other hand, its effect on the Gaussian term \eref{vertex:freeaction}
is to transform
\begin{equation}
  S_2[\R] \mapsto \int \measure{t}\,\dn{3}{x}\; \left[
  \frac{a^3}{\cs^2} \epsilon [\dot{\R}_n^2 + 2 \dot{\R}_n \dot{F}(\R_n)] -
  a \epsilon (\partial \R_n^2 + 2 \partial \R_n \partial F ) \right] .
\end{equation}
After integrating by parts, this is equivalent to
\begin{equation}
  S_2[\R] \mapsto \int S_2[\R_n] - 2 \int \measure{t}\,\dn{3}{x}\; F a
  \left.\frac{\delta L}{\delta \R}\right|_1 .
\end{equation}
According to this general argument,
we need to make a field redefinition 
$\R \mapsto \R_n + \frac{1}{4} \mathcal{F}$ in order 
to remove the terms arising in Eq. \eref{threept:redef}. 
Once this redefinition has been made, 
we can calculate the three-point function corresponding 
to the vertex \eref{threept:reduced}--\eref{threept:redef} 
by working only with Eq. \eref{threept:reduced} 
rewritten in terms of $\R_n$.

In the following section, we employ \eref{threept:reduced}
to calculate the three-point function 
$\langle \R(\vect{k}_1) \R(\vect{k}_2) \R(\vect{k}_3) \rangle$ 
of the fluctuation.

\section{Path integral formalism for the three-point function}
\label{sec:formalism}
The three-point function we wish to calculate is
$\langle \R(\vect{k}_1)\R(\vect{k}_2)\R(\vect{k}_3)\rangle$, which measures
correlations produced by the vertex
\eref{threept:reduced}--\eref{threept:redef}.  In this section we assume that
the field has been redefined to remove terms proportional to the equation of
motion, but for clarity in formulas we drop the subscript `$n$'.  This
expectation value is to be taken in the interacting vacuum $\ket{\Omega}$ of
the theory, that is,
\begin{equation}
  \langle \R(\vect{k}_1)\R(\vect{k}_2)\R(\vect{k}_3) \rangle =
  \bra{\Omega} \R(\vect{k}_1) \R(\vect{k}_2)
  \R(\vect{k}_3) \ket{\Omega} .
\end{equation}

\subsection{The interacting vacuum}
\label{sec:vacuum}
In order to assist the comparison of our formulae with those of 
\cite{maldacena-nongaussian,bartolo-matarrese-review}, we begin 
by briefly reviewing the standard construction of the 
interacting vacuum, following \cite{peskin-schroeder}. 
The quantization of a free field such as \eref{vertex:wavefn}, 
corresponding to the Gaussian action \eref{vertex:freeaction}, 
proceeds by writing
\begin{equation}
  \label{threept:freefield}
  \hat{\R}(t,\vect{k}) = \R_{\mathrm{cl}}(t,\vect{k}) \hat{a}(\vect{k}) +
  \R_{\mathrm{cl}}^\ast(t,\vect{k}) \hat{a}^\dag(\vect{k}) ,
\end{equation}
where $\hat{a}(\vect{k})$, $\hat{a}^\dag(\vect{k})$ are annihilation and
creation operators in the usual fashion, 
and $\R_{\mathrm{cl}}$ and $\R_{\mathrm{cl}}^\ast$ represent 
a classical solution of the field equation, such as 
that given in Eq. \eref{vertex:wavefn}, and its complex conjugate.  
The free vacuum $\ket{0}$ is constructed so that 
$\hat{a}(\vect{k})\ket{0} = 0$ for all $\vect{k}$.  
After introducing a self-interaction, 
such as the $\R^3$ vertex \eref{vertex:final}, 
the mode functions of $\R$ can no longer be calculated exactly 
because the interaction mixes Fourier modes. This implies that  
a construction such as that given in Eq. \eref{threept:freefield} 
is no longer possible.  

As a result, we must construct the 
interacting vacuum in a different manner.  One approach 
is to begin with \eref{threept:freefield} 
when the theory is approximately non-interacting, and evolve it 
forward according to the familiar Heisenberg rule
\begin{equation}
  \hat{\R}(t,\vect{k}) = \e{\imag H (t-t_0)} \hat{\R}(t_0,\vect{k})
  \e{-\imag H (t-t_0)},
\end{equation}
where $t_0$ is a fixed fiducial time at which \eref{threept:freefield} 
was constructed, and $H$ is the Hamiltonian.  
For example, the principle of cluster decomposition usually  
means that we can construct \eref{threept:freefield} 
as a solution of the Gaussian theory at asymptotic past infinity.  
In the cosmological context, this is when the mode corresponding to 
$\vect{k}$ is deep inside the horizon.

The Hamiltonian $H$ can be split into a piece $H_0$, corresponding to the 
Gaussian action \eref{vertex:freeaction}, and a piece $H_I$, 
corresponding to the self-interaction \eref{vertex:final}.  
It is then straightforward to verify that the quantum operator 
$\hat{\R}$ satisfies
\begin{equation}
  \label{threept:heisenberg}
  \hat{\R}(t,\vect{k}) = U^\dag(t,t_0) \hat{\R}_I(t,\vect{k}) U(t,t_0) ,
\end{equation}
where the interaction-picture field $\hat{\R}_I$ 
is a solution of the free field theory and 
the time-evolution operator $U$ is defined by
\begin{equation}
  U = \mathrm{T} \exp \left( - \imag \int_{t_0}^t \measure{\zeta} \;
  H_I(\zeta) \right) ,
\end{equation}
where $\mathrm{T}$ represents the time-ordering symbol.  
The interacting vacuum $\ket{\Omega}$ should be destroyed by annihilation 
operators corresponding to the full interacting theory, 
not the operators $a(\vect{k})$ of the Gaussian theory.  
To obtain $\ket{\Omega}$, one evolves $\ket{0}$ for some time $T$,
such that 
\begin{equation}
  \label{threept:vaca}
  \e{-\imag H T} \ket{0} = \e{-\imag E_0 T} \ket{\Omega} \braket{\Omega}{0} +
  \sum_{n \neq 0} \e{-\imag E_n T}
  \ket{n}\braket{n}{0} ,
\end{equation}
where $E_0 = \bra{\Omega} H \ket{\Omega}$, and the $\{ E_n \}$ 
are the spectrum of $H$.  Since $\ket{\Omega}$ is the vacuum of the theory, 
it follows that $E_0 < E_n$ for any $n$, so a slight rotation of  
$T$ into an imaginary direction, 
$T \rightarrow \infty(1 - \imag \delta)$, implies that 
all terms from the sum over $n \neq 0$ 
become exponentially small when compared to the term 
involving $\ket{\Omega}$. It follows that $\ket{\Omega}$ can be written
\begin{equation}
  \label{threept:vacb}
  \ket{\Omega} = \lim_{T \rightarrow \infty(1 - \imag \delta)}
  \frac{1}{\e{-\imag E_0 T} \braket{\Omega}{0}}
  \e{-\imag H T} \ket{0} .
\end{equation}
Combining this expression with the expression for the Heisenberg field, 
Eq. \eref{threept:heisenberg}, implies 
that one can compute the correlation functions of the interacting theory 
according to the rule
\begin{equation}
  \label{threept:correlate}
  \fl
  \bra{\Omega} \mathrm{T} \R(x) \cdots \R(y) \ket{\Omega} =
  \lim_{T \rightarrow \infty (1- \imag \delta)}
  \frac{\bra{0} \mathrm{T} \R_I(x) \cdots \R_I(y)
  \exp \left( - \imag \int_{-T}^T \measure{t} \, H_I(t) \right) \ket{0}}
  {\bra{0} \exp \left( - \imag \int_{-T}^T \measure{t} \, H_I(t) \right)
  \ket{0}} .
\end{equation}

\subsection{Tree-level three-point functions}
\label{sec:pathint}
Depending on the details of how $\ket{\Omega}$ relates to $\ket{0}$, 
one may wish to choose a more general contour of integration 
in \eref{threept:correlate} to project out the true interacting vacuum.  
Using this prescription, we can calculate the three-point 
function of the interacting $\R^3$ theory from the usual 
path integral rule:
\begin{equation}
  \label{tree:threept}
  \bra{\Omega} \R(t)^3 \ket{\Omega} = \frac{\int \fmeasure{\R} \; \R(t)^3 \,
  \exp(\imag \int_{\mathcal{C}} L)}
  {\int \fmeasure{\R} \; \exp(\imag \int_{\mathcal{C}} L)} ,
\end{equation}
where $\mathcal{C}$ is a contour of integration, chosen to select 
the interacting vacuum as described above, and 
$S = \int_{\mathcal{C}} L$ is the action.  
For the present theory this consists of the Gaussian 
piece \eref{vertex:freeaction}, which we write as $S_2$, 
and the three-field vertex \eref{threept:reduced}--\eref{threept:redef}, 
which we write as $S_3$.  
In principle, we should include diagrams containing  
an arbitrary number of loops when evaluating \eref{tree:threept}, 
but because $S_3$ is one order higher in slow-roll than $S_2$, 
the loop expansion is effectively the same as the expansion in slow-roll.  
As a result, if we are only interested in the leading-order slow-roll 
dependence, we need only calculate to tree level.

After expanding the interaction part of $\e{\imag S}$ as a 
series in the usual manner, and recalling that the integration of 
any odd number of $\R$'s against a Gaussian measure is identically zero, 
we find that the tree-level three-point function is given by 
\begin{equation}
  \label{pathint:expr}
  \bra{\Omega} \R(t)^3 \ket{\Omega}_{\mathrm{tree}} =
  \imag \frac{\int \fmeasure{\R} \; \R(t)^3 ( \int_{\mathcal{C}} L_3 )
  \exp(\imag \int_{\mathcal{C}} L_2)}{\int \fmeasure{\R} \;
  \exp(\imag \int_{\mathcal{C}} L_2)} .
\end{equation}

As an explicit example, let us 
consider a vertex of the form 
$L_3 = g \dot{\R}^2 \partial^{-2} \dot{\R}$ 
(where $g$ is some coupling constant) which arises 
in \eref{threept:vertex} after going to conformal time, 
and which we will need in the following section.  
In this case, one finds, after performing 
the functional integrations, 
that the three-point function can be written 
as $\langle \R(\tau_1,\vect{y}_1)\R(\tau_2,\vect{y}_2)\R(\tau_3,\vect{y}_3)
\rangle = \imag g \int_{\mathcal{C}} \measure{\eta}\,\dn{3}{x}\; \mathcal{G}$,
where the integrand $\mathcal{G}$ satisfies
\begin{equation}
  \fl
  \mathcal{G} = \sum_{\mathrm{pairings}}
  \deriv{}{\eta} \langle \R(\tau_1,\vect{y}_1)\R(\eta,\vect{x}) \rangle
  \deriv{}{\eta} \langle \R(\tau_2,\vect{y}_1)\R(\eta,\vect{x}) \rangle
  \partial^{-2}_{\vect{x}} \deriv{}{\eta} \langle \R(\tau_3,\vect{y}_3)
  \R(\eta,\vect{x}) \rangle
\end{equation}
and the sum is over all ways of pairing 
$\R(\tau_1,\vect{y}_1)$, $\R(\tau_2,\vect{y}_2)$ and 
$\R(\tau_3,\vect{y}_3)$ with fields in $L_3$.  
Notice that in principle, the path integral sums over 
\emph{all} pairings of the $\R$'s among themselves, including pairings 
of external fields with fields at the vertex, and pairings 
of external fields among themselves and vertex fields among themselves.  
However, there is no contribution from pairings where two external fields 
or two internal fields are paired, because the resulting amplitude 
is proportional to the tadpole for an $\R$ to emerge from the vacuum.  
Since we are assuming that the vacuum is stable, 
all such amplitudes vanish.  
This prescription is equivalent to dealing with 
only the connected three-point function.

After going over to Fourier space and using \eref{vertex:wavefn} 
for the propagator, one finds that 
\begin{eqnarray}
  \nonumber
  \fl
  \langle \R(\vect{k}_1) \R(\vect{k}_2) \R(\vect{k}_3) \rangle 
= -\imag g (2\pi)^3 \delta(\sum_i \vect{k}_i)
  \left(\frac{H^2}{4\epsilon\cs}\right)^3  \left(
  \frac{(1+\imag k \cs \tau)^3}{\prod_i k_i^3}\int_{-\infty}^{\tau} 
  \measure{\eta}
  \; k_1^2 k_2^2 \eta^2 \, \e{\imag K \cs(\eta-\tau)} \right. \nonumber \\
  \left. \mbox{} +
  \frac{(1-\imag k \cs \tau)^3}{\prod_i k_i^3}\int_{\tau}^0 \measure{\eta}
  \; k_1^2 k_2^2 \eta^2 \, \e{-\imag K \cs(\eta-\tau)} \right) 
  + \mbox{perms}  ,
  \label{pathint:aux}
\end{eqnarray}
where we have evaluated the correlation function at 
equal times $\tau_1 = \tau_2 = \tau_3 = \tau$ 
and $K = k_1 + k_2 + k_3$ is the total momentum.  
The mode is deep inside the horizon in the far past 
and oscillates rapidly.  In this r\'{e}gime, 
there is no contribution to the integral once 
we have rotated to Euclidean space.  In the far future, 
$\R$ tends to a constant and there is also no contribution.  
The dominant behaviour is determined by the modes' characteristics around 
horizon crossing.  Therefore, even though the propagator 
\eref{vertex:wavefn} is only valid under the assumption that 
$u$, $\epsilon$ and $\eta$ are small, 
Eq. \eref{pathint:aux} will still be a good estimate 
if the slow-roll parameters are sufficiently small 
around horizon crossing  and $H$, $\epsilon$ 
and $g$ are evaluated when these $k$-modes crossed the horizon. 
Taking these considerations into account, we arrive at the simple formula
\begin{equation}
  \label{pathint:vertex}
  \fl
  \langle \R(\vect{k}_1) \R(\vect{k}_2) \R(\vect{k}_3 \rangle =
  - \imag g (2\pi)^3 \delta(\sum_i \vect{k}_i)
  \left( \frac{H^2}{4\epsilon\cs}\right)^3 \frac{1}{\prod_i k_i^3}
  \int_{-\infty}^0 \measure{\eta}
  \; k_1^2 k_2^2 \eta^2 \, \e{\imag K \cs\eta} + \mbox{c.c.} ,
\end{equation}
plus permutations, where `c.c.' denotes the complex conjugate of the preceding
term, and we have taken the asymptotic value by sending $\tau \rightarrow 0$.

\section{The scalar non-gaussianity}
\label{sec:calc}
\subsection{The three-point function}
The interaction vertex (\ref{threept:reduced})
contains three terms. Two of these, proportional to 
$\dot{\R}^3$ and $\dot{\R}\partial^2 \R \partial^{-2}\dot{\R}$, 
are new and arise because the effective 
speed of sound $\cs \ne 1$. They vanish in the limit where 
$\cs =1$. The third, 
proportional to the $\dot{\R}^2 \partial^{-2}\dot{\R}$ interaction, 
generalizes the corresponding term derived in \cite{maldacena-nongaussian}
to the case where $u \neq 0$.  
At the tree-level approximation, the contribution that each of these 
interactions makes to the three-point 
correlation function is expressed in Eq. \eref{pathint:expr}.  
We now proceed to evaluate each of these contributions in turn.

\begin{itemize}
\item \textsl{$\dot{\R}^3$ interaction}.
Using the propagator \eref{vertex:wavefn}, and evaluating this 
vertex in a similar way to that outlined above, we find that 
this term gives a contribution
\begin{equation}
  \fl
  - \imag(2\pi)^3 \delta(\sum_i \vect{k}_i)\frac{H^4}{3 \cdot 2^5 \epsilon^2}
  \frac{u+\epsilon s/3\epsilon_X}{\prod_i k_i^3} \int_{\mathcal{C}}
  k_1^2 k_2^2 k_3^2 \eta^2 \, \measure{\eta} \, \e{\imag K \cs \eta}
  + \mbox{perms} + \mbox{c.c.}
\end{equation}
Since there are no poles, there is no obstacle to 
rotating the contour of integration so that it lies 
along the imaginary axis.  This can only be done in one direction, 
the direction compatible with the infinitesimal displacement 
$T \rightarrow \infty(1-\imag \delta)$, as described above.  
This means that we must rotate $\mathcal{C}$ 
clockwise onto $(\infty,0)\imag$, thereby yielding 
a total contribution equal to
\begin{equation}
  \fl
  -(2\pi)^3 \delta(\sum_i \vect{k}_i) \frac{H^4}{3\cdot 2^4 \epsilon^2}
  \frac{u + \epsilon s / 3 \epsilon_X}{\prod_i k_i^3}
  \frac{k_1^2 k_2^2 k_3^2}{K^3} + \mbox{perms} + \mbox{c.c.}
\end{equation}

\item \textsl{$\dot{\R}^2 \partial^{-2}\dot{\R}$ interaction}.
The second type of interaction contained in the vertex is the $u \neq 0$ 
generalization of the equivalent term that arises in the case of 
canonical inflation.  
This is the vertex we calculated in detail in 
\eref{pathint:vertex}, with $g = 4a^5 \epsilon (2u + \epsilon) H$.  
The contribution is therefore given by 
\begin{equation}
  \fl
  - \imag(2\pi)^3 \delta(\sum_i \vect{k}_i) \frac{H^7}{2^4 \epsilon^2} 
  \frac{1}{\prod_i k_i^3} (2u + \epsilon)
  \int_{\mathcal{C}} a^3 \eta^3 k_1^2 k_2^2 \, \measure{\eta} \,
  \e{\imag K \cs \eta} + \mbox{perms} + \mbox{c.c.}
\end{equation}
Since $a \eta = -H^{-1}$ during inflation (to leading order in slow-roll), 
all $\eta$ terms drop out of the integral except those arising 
in the exponential.  
After rotating the $\eta$-integral to the imaginary axis as in the 
previous interaction, we find that this contribution is given by 
\begin{equation}
  \fl
  (2\pi)^3 \delta(\sum_i \vect{k}_i) \frac{H^4}{2^4 \epsilon^2}
  \frac{1}{\prod_i k_i^3} (2u + \epsilon)
  \frac{k_1^2 k_2^2}{K} + \mbox{perms} + \mbox{c.c.}
\end{equation}

\item \textsl{$\dot{\R} \partial^2 \R \partial^{-2}\dot{\R}$ interaction}.
The remaining new piece in the vertex 
involves the term $\dot{\R}\partial^2 \R \partial^{-2}\dot{\R}$. This 
yields a correlation of the form
\begin{equation}
  \fl
  - \imag (2\pi)^3 \delta(\sum_i \vect{k}_i) \frac{H^6}{2^4 \epsilon^2}
  \frac{u}{\prod_i k_i^3} \int_{\mathcal{C}}
  a^2 \eta^2 k_1^2 k_2^2(1-\imag k_2 \cs \eta) \e{\imag K \cs \eta}
  + \mbox{perms} + \mbox{c.c.},
\end{equation}
with our usual conventions for $K$.  
After rotation of the contour $\mathcal{C}$, we find that this contribution 
reduces to 
\begin{equation}
  \fl
  - (2\pi)^3 \delta(\sum_i \vect{k}_i) \frac{H^4}{2^4 \epsilon^2}
  \frac{u}{\prod_i k_i^3} \left(
  \frac{k_1^2 k_2^2}{K} + \frac{k_1^2 k_2^3}{K^2} + \mbox{perms} + \mbox{c.c.}
  \right) .
\end{equation}
\end{itemize}

It is now necessary to take into account the 
contributions to the correlation function 
which enter via the field redefinition that was introduced to 
remove the terms in Eq. \eref{vertex:redef}. There are effectively 
two redefinitions, as summarized in Eq. (\ref{defF}), and 
we consider each of these in turn. 

\begin{itemize}
\item \textsl{The redefinition $\R \mapsto \R_n + 
(\eta - u - \epsilon)\R_n^2/4$}.
Under a field redefinition of the form 
$\R \mapsto \R_n + q \R_n^2$, Wick's theorem guarantees that
\begin{equation}
  \fl
  \langle \R(x)\R(y)\R(z) \rangle = q \langle \R(x)\R(y) \rangle\langle
  \R(x)\R(z) \rangle + \mbox{perms}.
\end{equation}
Any correlation function of the form 
$\langle \R(x)\R(y) \rangle\langle \R(x)\R(z) \rangle$ can 
be written in Fourier form by taking appropriate 
transforms in $x$, $y$ and $z$.  After making the necessary 
transformations and integrating out the $\delta$-functions 
\eref{vertex:twopt}, 
we obtain the product of two copies of the spectrum,
\begin{equation}
  \fl
  \frac{1}{4}(\eta - u -\epsilon) \frac{H^4}{16 \epsilon^2} \int
  \frac{\dn{3}{k_1} \, \dn{3}{k_2}}{(2\pi)^6} \; \e{\imag \vect{x}\cdot
  (\vect{k_1}+\vect{k}_2)}
  \e{-\imag\vect{y}\cdot\vect{k}_1}\e{-\imag\vect{z}\cdot\vect{k}_2}
  \frac{1}{k_1^3 k_2^3} +
  \mbox{perms},
\end{equation}
In order to express this result 
in a more familiar form, 
we introduce an auxiliary integration to represent 
the combination $\vect{k}_1 + \vect{k}_2$, 
constrained by a $\delta$-function, such that the contribution is 
equivalent to
\begin{equation}
  \fl
  \label{threept:qredef}
  \frac{1}{4}(\eta - u - \epsilon) \frac{H^4}{2^4 \epsilon^2} (2\pi)^3
  \int \frac{\dn{3}{k_1}\,\dn{3}{k_2}\,\dn{3}{k_3}}{(2\pi)^9} \;
  \delta(\sum_i \vect{k}_i)
  \e{\imag (\vect{k}_3\cdot\vect{x}+\vect{k}_1\cdot\vect{y}+
  \vect{k}_2\cdot\vect{z})}\frac{1}{k_1^3 k_2^3} +
  \mbox{perms} .
\end{equation}
In this form, the $k$-integrals and appropriate factors of 
$2\pi$ can be stripped off to yield 
the correct $k$-space contribution to the correlation function.

\item \textsl{The redefinition $\R \mapsto \R_n + (\epsilon/2)
\partial^{-2}(\R_n \partial^2 \R_n)$.}
In this case we obtain contributions to the three-point correlation function of
the form
\begin{equation}
  \langle \R(x)\R(y)\R(z) \rangle = \frac{\epsilon}{2} \partial_x^{-2} \left(
  \langle \R(x)\R(y) \partial_x^2
  \langle \R(x)\R(z) \rangle \right) .
\end{equation}
Introducing the Fourier transform of the two-point functions, as outlined above, 
and integrating out $\delta$-functions, it follows that 
this term must be identical to 
\begin{equation}
  -\frac{H^4}{32\epsilon} \partial_x^{-2} \int \frac{\dn{3}{k_1}\,
  \dn{3}{k_2}}{(2\pi)^6}
  \e{\imag\vect{x}\cdot(\vect{k}_1+\vect{k}_2)}\e{\imag\vect{y}\cdot\vect{k}_1}
  \e{\imag\vect{z}\cdot\vect{k}_2}
  \frac{1}{k_1 k_3^3} + \mbox{perms} .
\end{equation}
As before, by introducing an auxiliary constrained 
integration to represent $\vect{k}_1 + \vect{k}_2$, 
this can be written in the form 
\begin{equation}
  (2\pi)^3 \delta(\sum_i \vect{k}_i) \frac{H^4}{2 \cdot 2^4 \epsilon}
  \frac{1}{k_1 k_2^2 k_3^3} + \mbox{perms} ,
\end{equation}
\end{itemize}
after $k$-space integrations and factors of $2\pi$ have been removed.

Bringing all of these separate contributions together, we arrive at 
an expression for the three-point scalar correlation function: 
\begin{equation}
  \label{threept:fn}
  \langle \R(\vect{k}_1) \R(\vect{k}_2) \R(\vect{k}_3) \rangle =
  (2\pi)^3 \delta(\sum_i \vect{k}_i)
  \frac{H^4}{2^4 \epsilon^2} \frac{1}{\prod_i k_i^3} \mathcal{A} ,
\end{equation}
where the $k$-dependence is determined by the function $\mathcal{A}$:
\begin{eqnarray}
  \label{threept:afn}
  \fl
  \mathcal{A} = \frac{4}{K}(u + \epsilon) \sum_{i>j} k_i^2 k_j^2
 - \frac{4}{K^3}\left(u + \frac{\epsilon}{\epsilon_X}
  \frac{s}{3} \right) k_1^2 k_2^2
  k_3^2
\nonumber \\
 - \frac{2u}{K^2} \sum_{i \neq j} k_i^2 k_j^3 + \frac{1}{2}(\eta - u - \epsilon)
  \sum_i k_i^3 + \frac{\epsilon}{2} \sum_{i \neq j} k_i k_j^2 .
\end{eqnarray}
In evaluating this expression, 
we have assumed that $k_1 \sim k_2 \sim k_3$, so that the 
epoch of horizon crossing for each of the $k$-modes is comparable.  
Eqs.~\eref{threept:fn}--\eref{threept:afn} 
constitute the principal result of this paper.

\subsection{Consistency conditions}
\label{sec:consistency}

The full three-point function \eref{threept:fn}--\eref{threept:afn}
is rather complicated, and 
one would like to have an independent check of its validity.  
This may be achieved by relating the three-point function 
to the spectral index of the two-point function in an appropriate limit, as
discussed in \cite{maldacena-nongaussian,creminelli-zaldarriaga,gruzinov}.
We begin by briefly reviewing the 
argument of \cite{maldacena-nongaussian} within the context of models 
where $\cs \neq 1$.  
The aim is to write down a long-wavelength \emph{gravitational} 
consistency condition between the two- and three-point correlation 
functions. In the limit where one momentum, 
say $\vect{k}_3$, is much smaller in 
magnitude than the other two, such that 
$k_3 \ll k_1 \sim k_2$, the remaining momenta 
$\vect{k}_1$ and $\vect{k}_2$ become approximately equal and opposite.  
From the point of view of physical perturbations, 
this means that the wavelength of the $\vect{k}_3$ mode 
has been made arbitrarily long.  
A mode such as this, which is deep in the infrared and far 
outside any individual observer's horizon, 
is effectively just a renormalization of the background theory, which
in this case corresponds to a rescaling of the spatial coordinates 
$x^i$ by a factor $\e{\R_3} \sim (1 + \R_3)$, 
i.e., $\delta x = \R_3 x$, where $\R_3$ is the amplitude of the 
$\vect{k}_3$ mode.  
Such a rescaling implies that the wavenumber of any mode still 
inside the horizon must be compressed by a corresponding amount 
$\delta k = - \R_3 k$ and 
this compression results in a change in the epoch of horizon exit, 
$k \cs = a H $, such that 
\begin{equation}
\delta k \cdot \cs = \delta a \cdot H .
\end{equation}

If we use the standard relation $\delta a = a H \delta t$, this implies that 
$\delta t = - \R_3 / H$ {\em independently} 
of the value of $\cs$.  As a result, 
we expect the three-point function in this limit 
to express how the $\vect{k}_3$ mode correlates 
with the change of the two-point function of $\vect{k}_1$ 
and $\vect{k}_2$ induced by differences in the epochs of horizon exit,
i.e., 
\begin{equation}
  \label{threept:consistency}
  \langle \R(\vect{k}_1) \R(\vect{k}_2) \R(\vect{k}_3) \rangle \rightarrow
  (2\pi)^3 \delta(\sum_i \vect{k}_i) (1-n) P(k_1) P(k_3)  ,
\end{equation}
where $P(k)$ was defined in \eref{vertex:twopt}, 
and $n$ is the spectral tilt \eref{vertex:tilt}.  
This consistency equation was rederived on kinematical grounds 
in \cite{creminelli-zaldarriaga}, where it was emphasized that 
\eref{threept:consistency} holds independently of the details 
of the matter Lagrangian.  Therefore, one should expect 
\eref{threept:fn}--\eref{threept:afn} to obey 
\eref{threept:consistency} as $|\vect{k}_3|$ is sent to zero.

In fact, it is straightforward to verify 
that the momentum sums in \eref{threept:afn} behave according to
\begin{equation}
  \fl
  \sum_{i > j} \frac{k_i^2 k_j^2}{K} \sim \frac{1}{2} k_2^3, \qquad
  \sum_{i \neq j} \frac{k_i^2 k_j^3}{K^2} \sim \frac{1}{2} k_2^3, \qquad
  \sum_{i} k_i^3 \sim 2 k_2^3, \qquad 
  \sum_{i \neq j} k_i k_j^2 \sim 2 k_2^3 ,
\end{equation}
whereas $k_1^2 k_2^2 k_3^2 \sim 0$.  This behaviour is sufficient to show that
\begin{equation}
  \label{threept:mcz}
  \langle \R(\vect{k}_1 \R(\vect{k}_2) \R(\vect{k}_3) \rangle \rightarrow
  (2\pi)^3 \delta(\sum_i \vect{k}_i)
  \frac{H^4}{2^4\epsilon^2} \frac{1}{k_1^3 k_3^3} (2\epsilon + \eta) ,
\end{equation}
which reduces to \eref{threept:consistency} when the constant 
of proportionality is rewritten in terms of the scalar spectral 
index.

An estimate of the level of non-gaussianity generated purely 
by the scalar field, rather than its coupling to gravity,  
can be derived by specifying $\epsilon = \eta = 0$.  
In this r\'{e}gime, the $k$-dependence of (\ref{threept:afn}) reduces to 
\begin{equation}
  \label{threept:matterfn}
  \mathcal{A}|_{\epsilon=\eta=0} =
  4u \sum_{i > j} \frac{k_i^2 k_j^2}{K} - 4u
  \frac{k_1^2 k_2^2 k_3^2}{K^3} - 2 u \sum_{i \neq j} \frac{k_i^2 k_j^3}{K^2} -
  \frac{u}{2} \sum_i k_i^3 ,
\end{equation}
and, in particular, $\mathcal{A}|_{\epsilon=\eta=0} \rightarrow 0$ 
as $k_3 \rightarrow 0$.  The behaviour of $\mathcal{A}|_{\epsilon=\eta=0}$ in
this limit guarantees that \eref{threept:mcz} is independent of the speed of
sound.

\subsection{Observational considerations}
\label{sec:observe}
The level of accuracy currently achievable by observations is 
insufficient for a projection of the three-point function 
\eref{threept:fn}--\eref{threept:afn} onto the CMB sky 
to be measured directly \cite{wang-kamionkowski,babich-creminelli}. 
Instead, observational limits are set on a 
parameter $\fnl$ \cite{komatsu-spergel,verde-wang}, which measures the
departure from Gaussianity of the gravitational potential.  This
potential \cite{sachs-wolfe},
together with other detailed physics at the last-scattering
surface, controls the character of the CMB anisotropies we measure
at the present epoch.
These effects must be taken into consideration when calculating
the non-linearity $\fnl$ that a given model of inflation will imprint in
the CMB \cite{bartolo-matarrese-fnl,bartolo-matarrese-inflation}.
Present-day observations impose a conservative upper 
bound of $|\fnl| < 100$ \cite{wmap-gaussian}, 
whereas a non-gaussian signal exceeding $\fnl \geq 5$ 
should be detectable from forthcoming observations of the 
Planck satellite \cite{komatsu-spergel}.

Quite generally, one can write $\fnl$ as the sum of a superhorizon piece,
which is is produced by gravitational evolution of the superhorizon
perturbations after inflation has ended \cite{bartolo-matarrese-inflation},
and a primordial piece, $\fnl^\R$,
related to the non-linearity in the curvature
perturbation $\R$, which should properly be regarded as a kind of initial
condition.  This initial condition is most conveniently characterized by
writing
\begin{equation}
  \label{obs:fnl}
  \R(x) = \R_{\mathrm{g}}(x) - \frac{3}{5} \fnl^\R \star (\R_{\mathrm{g}}(x)^2 -
  \langle \R_{\mathrm{g}}^2 \rangle),
\end{equation}
where $\R_{\mathrm{g}}(x)$ is a Gaussian field, 
and a convolution product $\star$ is involved since 
in general $\fnl^\R$ may be a function of scale. 
The standard single-field inflationary scenario predicts 
$\fnl^\R \sim \Or(\epsilon)$ \cite{maldacena-nongaussian,acquaviva-bartolo},
whereas the
universal superhorizon contribution is $\sim \Or(1)$.  In consequence,
a positive 
detection of primordial non-gaussianities from WMAP or Planck
would represent one possible way of ruling out such models.  
We will see that the prediction for non-canonical inflation is similar.

The field \eref{obs:fnl} is just a particular example of a redefinition 
of the form \eref{threept:qredef}, so it is straightforward to express 
$\fnl$ in terms of $\mathcal{A}$:
\begin{equation}
  \label{obs:fnlvalue}
  \fnl^\R = - \frac{5}{6} \frac{\mathcal{A}}{\sum_i k_i^3} .
\end{equation}
Eq.~\eref{obs:fnlvalue} can be written explicitly in terms of 
momentum sums if desired, but the result is complicated and 
ultimately not informative.  A better estimate of the size of 
any non-gaussianity produced in this model is given by evaluating 
$\fnl^\R$ on equilateral triangles, where $k_1 = k_2 = k_3$.  
It is a simple matter to calculate
\begin{eqnarray}
  \fl
  \left.\fnl^\R\right|_{\mathrm{equilateral}} = - \frac{275}{972} u + 
  \frac{10}{729} \frac{\epsilon}{\epsilon_X} s
  - \frac{55}{36}\epsilon - \frac{5}{12} \eta \nonumber \\
  \lo\approx
  - 0.28 u + 0.02 \frac{\epsilon}{\epsilon_X} s - 1.53 \epsilon - 0.42 \eta .
\end{eqnarray}

This value of $\fnl^\R$ agrees with the corresponding result of 
Maldacena \cite{maldacena-nongaussian} for standard inflation, 
for which $s=u=0$, and can be compared with predictions for similar 
quantities in alternative models \cite{alishahiha-silverstein,creminelli}.  
For example, Creminelli \cite{creminelli} has considered a model that 
in our notation corresponds to 
specifying
\begin{equation}
  P(X,\phi) = X/2 - V(\phi) + X^2/8M^4,
\end{equation}
where $M$ is some mass scale.  
As discussed in the introduction, 
$M$ should be associated with a large renormalization scale 
which cuts off the details of the physics in the far ultraviolet limit.  
Although gravity was neglected in the analysis of \cite{creminelli}, 
the scale $M$ is related to the speed of sound in such a way that 
\begin{equation}
  \cs^2 = \frac{2M^4 + X}{2M^4 + 3X} .
\end{equation}
Moreover, to leading order in $M^{-4}$, 
the slow-roll parameters $u$ and $s$ can be written as
\begin{eqnarray}
  u = - \frac{2X}{2M^4 + X} \sim - \frac{X}{M^4}
 \nonumber \\
  \frac{\epsilon}{\epsilon_X} s = \frac{12 X M^4}{(2M^4+X)(2M^4+3X)}
  \sim 3 \frac{X}{M^4} .
\end{eqnarray}

Thus, $u$ is small and slow-roll (in the generalized sense of 
Section~\ref{sec:sr}) is a good approximation if $M^4 \gg X$, or 
equivalently, if $|\dot{\phi}| \ll M^2$.  
In this limit, and setting $\epsilon = \eta = 0$ to decouple the 
contribution from gravity, we obtain a non-gaussianity of magnitude
\begin{equation}
  \fnl^\R = \frac{35}{108} \frac{\dot{\phi}^2}{M^4} \approx 
0.32 \frac{\dot{\phi}^2}{M^4} 
\end{equation}
in exact agreement with \cite{creminelli}.  
Although the non-gaussianities can be made arbitrarily 
large by sending $M^4$ to zero, it is important to note that
$u$ ceases to become small in this limit, and indeed
$\cs \rightarrow 1/\sqrt{3}$.  Therefore,
the slow-roll approximation will eventually break down 
well before the non-gaussianities have become large.

\section{Conclusions}
\label{sec:conclusions}
In this paper, we have derived the 
three-point correlation function of scalar fluctuations 
generated during an early inflationary epoch 
in the case where the inflaton Lagrangian is an arbitrary function of the 
field and its first derivative. This has applications to the 
$k$-inflationary scenario \cite{armendariz-picon,garriga-mukhanov} 
and inflation driven by a ghost condensate 
\cite{arkani-hamed-chang,arkani-hamed-creminelli}. 
More generally, it applies to any theory 
containing higher derivative operators 
which descends from a higher-dimensional supergravity or superstring 
compactification, or includes the effect of radiative corrections.  

Our computation reproduces previously derived results
\cite{maldacena-nongaussian,acquaviva-bartolo}
in the appropriate limit of a canonically coupled 
scalar field, and respects the consistency relation 
proposed on kinematical grounds by Maldacena 
\cite{maldacena-nongaussian} and Creminelli \& Zaldarriaga
\cite{creminelli-zaldarriaga}. 
In the latter case, the reduction of the three-point function 
\eref{threept:fn}--\eref{threept:afn} to the consistency condition 
\eref{threept:consistency} is non-trivial, 
since the factors of $u$ which are 
distributed over the various $k$-dependent terms must sum to zero 
as $k_3 \rightarrow 0$.  

This pattern of $k$-dependence in the three-point function 
is quite different to that of a slowly rolling 
field with a canonical kinetic term, 
but similar to results derived in previous attempts to go beyond 
canonical inflation \cite{creminelli,gruzinov}.  
In principle, this modification in angular dependence provides a sharp 
discriminant between these different classes of models.  
However, the overall scale of the non-gaussianity \eref{obs:fnlvalue} 
is small, being proportional to a sum of flow parameters
which are suppressed below unity when slow-roll is valid, as is
required for self-consistency of the calculation.
This implies that if the density fluctuations which seeded the
CMB anisotropies were generated during an 
epoch of slow-roll inflation of the type we have considered, the level of 
non-gaussianity will be small even in the presence of 
higher-derivative operators of the inflaton field.

If the slow-roll conditions described in Section~\ref{sec:sr} do not apply,
then the situation is less clear.  It is possible to have 
quasi-exponential inflation,
$\epsilon \ll 1$, even if $|u|$ or $|s|$ are large.
In this case, a relatively large measure 
of non-gaussianity could be produced during inflation by reducing the 
speed of sound, through a suitable choice for the functional form of 
$P(X,\phi)$.  This possibility may become attractive if 
forthcoming observations
indicate the presence of an appreciable 
measure of primordial non-gaussianity in the
CMB.  However, whenever $u$ is large 
(so $\cs$ departs from the speed of light by
a significant amount), the slow-roll approximation 
begins to break down and the
calculation we have described is no longer appropriate.  
In this limit, we
expect the reduced speed of sound to cause the spectral index 
to deviate from unity. Consequently, a larger non-gaussian 
signal will typically be accompanied by a larger deviation 
from a Harrison--Zel'dovich spectrum.

To make further progress, it will be 
necessary to obtain the solution of the Mukhanov equation when 
one or more of the slow-roll parameters undergoes significant 
variation. Some progress
has recently been made in this direction \cite{stewart-gong,wei-cai-wang},
but in any event, the effective field theory 
for the inflaton will probably cease to be valid before $u$ can be made very
large \cite{creminelli}.

\ackn
DS is supported by PPARC.

\section*{\refname}

\providecommand{\href}[2]{#2}\begingroup\raggedright\endgroup

\end{document}